\documentclass[aps,prb,preprint,groupedaddress]{revtex4-1}
\usepackage{amsmath,amsfonts,amssymb}
\usepackage{graphicx,url,hyperref}
\usepackage{tabularx}
\usepackage{multirow}
\usepackage{color}
\usepackage{makecell}
\usepackage{caption}
\usepackage{subcaption}
\usepackage{bm}
\newcommand{\mr}[1]{\mathrm{#1}}
\begin{document}

\title{Simultaneous Learning of Several Materials Properties from Incomplete Databases with Multi-Task SISSO}

\author{Runhai Ouyang}
\email{ouyang@fhi-berlin.mpg.de}
\affiliation{Fritz-Haber-Institut der Max-Planck-Gesellschaft, 14195 Berlin-Dahlem, Germany}
\author{Emre Ahmetcik}
\affiliation{Fritz-Haber-Institut der Max-Planck-Gesellschaft, 14195 Berlin-Dahlem, Germany}
\author{Christian Carbogno}
\affiliation{Fritz-Haber-Institut der Max-Planck-Gesellschaft, 14195 Berlin-Dahlem, Germany}
\author{Matthias Scheffler}
\affiliation{Fritz-Haber-Institut der Max-Planck-Gesellschaft, 14195 Berlin-Dahlem, Germany}
\author{Luca M. Ghiringhelli}
\email{ghiringhelli@fhi-berlin.mpg.de}
\affiliation{Fritz-Haber-Institut der Max-Planck-Gesellschaft, 14195 Berlin-Dahlem, Germany}

\date{\today}

\begin{abstract}
The identification of descriptors of materials properties and functions that capture the underlying physical mechanisms is a critical goal in data-driven materials science.
Only such descriptors will enable a trustful and efficient scanning of materials spaces and possibly the discovery of new materials. 
Recently, the sure-independence screening and sparsifying operator (SISSO) has been introduced and was successfully applied to a number of materials-science problems. SISSO is a compressed-sensing based methodology yielding predictive models that are expressed in form of analytical formulas, built from simple physical properties. These formulas are systematically selected from an immense number (billions or more) of candidates. In this work, we describe a powerful extension of the methodology to a `multi-task learning' approach, which identifies a single descriptor capturing multiple target materials properties at the same time. This approach is specifically suited for a heterogeneous materials database with scarce or partial data, e.g., in which not all properties are reported for all materials in the training set. As showcase examples, we address the construction of materials-properties maps for the relative stability of octet-binary compounds, considering several crystal phases simultaneously, and the metal/insulator classification of binary materials distributed over many crystal-prototypes.
\end{abstract}


\maketitle

\section{Introduction}
The materials-genome initiative \cite{MGI} inspired the establishment of several high-throughput computational materials-science projects, leading to the creation of worldwide accessible materials databases~\cite{aflow, materialsproject.org, oqmd.org, cmr_repository}. 
In this context, the Novel Materials Discovery (NOMAD) Repository \& Archive is the biggest data base for input and output files of density-functional theory calculations for materials considering all important computer codes of the community \cite{LMG2016,Draxl2018,Draxl@Yip}. It plays synergistically together with other important data bases, in particular AFLOW \cite{aflow}, Materials Project \cite{materialsproject.org}, and  OQMD \cite{oqmd.org}.

This wealth of available data opens the era of the data-driven materials science \cite{Hey_Forth_Paradigm_Book_Microsoft,Draxl2018}, which is fueled by the computer-aided analysis of the data, in order to find patterns and trends otherwise invisible to the human eye. This, in turn, may lead to accelerate discoveries of new materials or phenomena.

A  key goal of materials science is to find materials with a high performance in several functions, e.g., stability and catalytic activity and selectivity for a very specific chemical reaction. It is important to realize that the number of materials that qualify is typically very small. However, the complexity and intricacy of the actuating processes is significant. 
Falling under the umbrella names of artificial intelligence or (big-)data analytics (terms that include data mining, machine/statistical learning, deep learning, compressed sensing, etc.), several methods have been developed and applied to existing materials-science data \cite{Bartok_PRL_2010,curtarolo:art85,Rajan_AnnRev2015,Mueller_MLMS_2016_review,Kim_ChemMat_2016,Lilienfeld2016,Takahashi_DalTrans2016,Ceriotti2017,Goldsmith_NJP_2017,Pham_jcp2018} in order to predict properties of interest. 

The $T=0$~K properties of materials are fully described by the many-body Hamiltonian, which is uniquely identified by its descriptors: the position and charges of the atomic nuclei $\{ R_{\mr{I}}, Z_{\mr{I}} \}$ and the number of electrons $N^e$. Although, in principle, these could be also descriptors for an artificial-intelligence algorithm, their connection with the materials properties and functions is too complicated, indirect, intricate. As a consequence, the description of processes ruling materials properties and functions requires to add as much domain knowledge to the artificial-intelligence step as available. Obviously, if not done with utmost care, this may well yield a biased and unreliable description. From the mentioned ``fundamental primary'' descriptors, $\{R_{\mr{I}}, Z_{\mr{I}}\}$ and $N^e$, it is also clear that there are two types of needed information: 1) the topology of the atomic structure and 2) the electronic/chemical property of the atoms. When geometry changes are not relevant (or trivial) the first aspect can be simplified or even neglected, and when changes in chemical bonding are nor relevant (or trivial), the second aspect can be simplified or even neglected. We will get back to these issues in the specific application examples discussed below.

Following the strategy introduced in Ref. \onlinecite{Luca2015}, the descriptor can be learned from the data, more precisely the best descriptor can be identified among a possibly immense set of candidates by exploiting a signal-analysis technique known as compressed sensing (CS) \cite{Candes2006,Donoho2006,Nelson2013,Luca2015,Luca2017}. 
SISSO\cite{SISSO} is a recently developed CS-based method, designed for identifying low-dimensional descriptors (a descriptor is defined as a vector of features, so that the number of features is the dimension of the descriptor) for material properties. It is an iterative scheme that combines the sure independence screening (SIS)\cite{SIS} scheme for dimensionality reduction of huge features space and the sparsifying operators for finding sparse solutions. SISSO improves the results over conventional CS methods such as the Linear Absolute Shrinkage and Selection Operator (LASSO\cite{LASSO}), or LASSO-based \cite{Luca2015,Luca2017} and greedy algorithms \cite{Tropp2007,OMP} when features are correlated, and can efficiently manage immense features spaces. SISSO has been already successfully applied to identifying descriptors for relevant materials-science properties\cite{SISSO,Bartel2018a,Bartel2018b}.

In this work, we introduce a learning scheme, termed multi-task (MT) SISSO, within the framework of the wider class of learning schemes known as multi-task learning (MTL)\cite{caruana97,MTL,Obozinski2006,Argyriou2008,Yin2018,Gong2013,Huang2015,Thung2018,Zhang2018}.
A {\em task} for a learning algorithm is the learning of a target property starting from a single input source (set of features). The learning of {\em multiple tasks} (or MTL) is an umbrella term that refers to\cite{Thung2018} $(i)$ the learning of multiple target properties using a single input source, {\em or} $(ii)$ the joint learning of a single target property using multiple input sources, {\em or} $(iii)$ a mixture of both. The key aspect is the parallel learning of multiple tasks, with the (sometimes implicit) assumption that the shared information among different tasks can lead to better learning performance if all the tasks are learned jointly, as compared to learning them independently. In other words, MTL assumes that the learning of one task can improve the learning of the other tasks \cite{Thung2018}. 
Though MTL has not yet been applied to materials-science problems so far, it has already been widely applied in other fields, such as in the handwriting recognition problem, self-driving automation system, computer vision, bioinformatics and health informatics, speech and language recognition, and more.\cite{caruana97,Obozinski2006,Thung2018,Zhang2018}

In order to clarify how the MTL concept can be applied in materials science, let us introduce the showcase examples that will be addressed in the following sections. Arguably one of the fundamental challenge in materials science is predicting the ground-state crystal structure of a material, given its chemical composition. In Refs. \onlinecite{Luca2015,Luca2017,SISSO}, models for predicting the relative stability or rock-salt vs zinc-blende structures for $AB$ octet binaries were learned via a LASSO-based and the SISSO algorithms. Learning models for the prediction of the relative stability of more than two crystal structures, given the same set of chemical formulas, can be cast into MTL. Each difference in energy between crystal structures is a {\em task} and the common input is the chemical formula and/or a list of properties of the atomic species listed in the chemical formula. The joint learning, in the SISSO framework, sets in when the same descriptor is imposed to be selected for all tasks. More specifically, SISSO identifies models in form of linear mappings between the descriptor $\bm{d}$ --- a vector of nonlinear functions of physical properties termed {\em primary features} --- and the property of interest $P = \bm{dc}$, where $\bm{c}$ is the vector of coefficients that maps $\bm{d}$ into $P$. If we now consider a set $\{ P^{(1)}, P^{(2)}, \ldots, P^{N^\mr{T}} \}$ of $N^\mr{T}$ properties  (e.g., the set of energy differences between crystal structures for the same chemical formula), the idea of MTL applied to SISSO is to find models $P^k = \bm{d}\cdot\bm{c}^k$ where the set of fitting coefficients $\{ \bm{c}^{(1)}, \bm{c}^{(2)}, \ldots, \bm{c}^{N^\mr{T}} \}$ maps {\em the same descriptor} $\bm{d}$ into the different properties $\{ P^{(1)}, P^{(2)}, \ldots, P^{N^\mr{T}} \}$. In section \ref{MT-SISSOcont-appl}, we will show the results of such learning. Besides the physical meaningfulness and Occam-razor-reminiscent elegance that a few mechanisms are ruling all energy differences (though with different relative importance), a great advantage of the MTL framework is to allow for a robust learning also when the training database (in this case, reference energy differences) is incomplete, i.e., for several chemical formulas only some of the energy differences are known. As we will show, MT-SISSO learns accurate predictive models also with high levels of incompleteness (e.g., when 50\% or more of the information is randomly missing).

A second setup where MT-SISSO is helpful is the learning of one common property of many materials belonging to physically different groups, e.g., they have different bonding characteristics and their ground-state crystal structure belong to different space groups. Obviously, in such situation one single predictive model is difficult to be found. 
This is the setup of our second showcase application (see section \ref{MT-SISSOclass-appl}) where the challenge is to find a model for predicting whether a material is a metal or nonmetal, with materials belonging to many different crystal-prototype classes. More specifically, we address the construction of two-dimensional maps where materials being metals or nonmetals are located in two non-overlapping convex regions. In MTL language, each map --- one for each crystal prototype --- is a {\em task} and the joint learning imposes that all maps share the same descriptor (in practice the same quantities on the axes). 
The metal/nonmetal classification challenge was already tackled with (single-task) SISSO in Ref. \onlinecite{SISSO}, but here, with an enlarged, heterogeneous  materials space (more crystal prototypes), only MT-SISSO is able to achieve an accurate description. Similarly to the previous example, one key feature of the use of MT-SISSO is the possibility to learn predictive models by omitting a significant amount of data from the training database. 

Before describing our showcase examples, in the following section we introduce the methodology and notation of MT-SISSO,

\section{Methodology}
\subsection{Single-task SISSO for continuous property} \label{SISSOcont}
In order to underline the analogies and crucial differences between single-task (ST) and MT-SISSO, we start with a brief recapitulation of the ST-SISSO algorithm. A detailed explanation of the SISSO algorithm is given in Ref. \onlinecite{SISSO} and a recommended hands-on tutorial is given in the online Python notebook\cite{CS_tutorial} at the {\em NOMAD Analytics Toolkit}\cite{Analytics-Toolkit} website.
The setup of ST-SISSO starts from a given set of materials with scalar-valued, continuous properties listed in a vector $\bm{P}$ (an element $P_i$ of $\bm{P}$ is the property of the $i$-th material) and a  --- typically huge --- list of $N^\mr{D}$ possible candidate features forming the {\em features space}. The projection of each $i$-material into the $j$-feature yields the $i,j$ component of the ``sensing matrix'' $\bm{D}$, having $N^\mr{M}$ rows and $N^\mr{D}$ columns, with $N^\mr{D} \gg N^\mr{M}$.
The solution of 
\begin{equation}
 \operatorname*{arg\,min}_{\bm{c}}\left( \lVert\bm{P}-\bm{Dc}\rVert^2_2 +\lambda\lVert{\bm{c}}\rVert_0 \right)
 \label{eq:SISSO0}
\end{equation}
where $\lVert{\bm{c}}\rVert_0$ is the $\ell_0$ norm of $\bm{c}$, i.e., the number of nonzero components of $\bm{c}$, gives the optimum $\Omega$-dimensional descriptor, i.e., the set of features singled out by the $\Omega$ non-zero components of the solution vector $\bm{c}$. The parameter $\lambda$  weights the relative importance of training accuracy vs dimensionality $\Omega$ (known as ``sparsity'' in the CS language).

The feature space $\bm{\Phi}_q$ is constructed by starting from a set of primary features $\bm{\Phi}_0$ and a set of unary and binary operators (such as $ +,-,\exp{},\sqrt{\phantom{0}}$, \ldots). The features are then iteratively combined with the operators, where at each iteration each feature (pair of features) is exhaustively combined with each unary (binary) operator, with the constraint that sums and differences are taken only among homogeneous quantities. The index $q$ in $\bm{\Phi}_q$ counts how many such iterations were performed. The primary features are typically physical properties of gas-phase atoms (e.g., ionization potential, radius of $s$ ot $p$ valence orbital, etc.) and {\em collective} properties of group of atoms (e.g., formation energy of dimers, volume of the unit cell in a given crystal structure, average coordination, etc.)\cite{SISSO}.
The features in $\bm{\Phi}_q$ are represented in terms of mathematical expressions. The evaluation of the $j$-th feature for all the $N^\mr{M}$ materials provides the $j$-th column in the sensing matrix $\bm{D}$. The properties of gas-phase atoms --- in short, {\em atomic properties} --- are ``repurposable'', in the sense that they can be used for many descriptor and model learning procedures. For easier reference and reusability, the atomic features used in this work and other related works \cite{Luca2015,Luca2017,SISSO,Bartel2018a} can be accessed on line at the {\em NOMAD Analytics Toolkit}. A tutorial\cite{PeriodicTable_tutorial} shows how to access these quantities and use them in a python notebook.

The algorithm for addressing Eq. \ref{eq:SISSO0} with ST-SISSO is: \\
$(i)$ SIS preliminary step. A subspace $\bm{S}_{1}$ is selected containing the $N^{\bm{S}}_{1}$ features having the largest linear correlation (largest absolute value of scalar product) with $\bm{P}$. The feature vector $\bm{d}_{1}$ --- the column of $\bm{D}$ with the largest correlation with $\bm{P}$ --- is the one-dimensional ($\Omega=1$) SISSO solution and also the exact 1D solution of Eq. \ref{eq:SISSO0}.\\
$(ii)$ Evaluation of the residual $\bm{\Delta}_{1} \equiv \bm{P}-\bm{d}_{1} c_{1}$, where the scalar $c_{1}=({\bm{d}_{1}}^T\bm{d}_{1})^{-1}{\bm{d}_{1}}^T\bm{P}$ is the least square solution of fitting $\bm{d}_{1}$ to $\bm{P}$. \\
$(iii)$ SIS step of iteration $\Omega>1$, which consists in selecting the subspace of the $N^{\bm{S}}_{\Omega}$ features with largest correlation with $\bm{\Delta}_{(\Omega-1)}$ and take the union of this subsets with $\bm{S}_{(\Omega-1)}$ to form $\bm{S}_\Omega$. \\
$(iv)$ SO step of iteration $\Omega>1$. Several SO strategies are possible; in this paper (as in Ref.\onlinecite{SISSO}), we adopt the so-called $\ell_0$ regularization, which finds the exact optimum solution within the subset $\bm{S}_\Omega$ selected by SIS. For all possible $\Omega$-tuples in $\bm{S}_\Omega$, it finds the one that gives the smallest $\ell_2$ (Euclidean) norm of the residual $\bm{\Delta}_\Omega \equiv \bm{P}-\bm{d}_\Omega \bm{c}_\Omega$, where $\bm{d}_\Omega$ is the matrix whose columns are the members of the considered $\Omega$-tuple and the vector $\bm{c}_\Omega=({\bm{d}_\Omega}^T\bm{d}_\Omega)^{-1}{\bm{d}_\Omega}^T\bm{P}$ is the least square solution of fitting $\bm{d}_{\Omega}$ to $\bm{P}$.
Points $(iii)$ and $(iv)$ are iterated until the stopping criterion is met. For instance, one stopping criterion (used in the application described in section section \ref{MT-SISSOcont-appl}) is that the $\ell_2$ norm of $\bm{\Delta}_\Omega$ is smaller than a prefixed threshold. The $\Omega$-dimensional descriptor identified by ST-SISSO is $\bm{d}_\Omega$ and the related predictive {\em model} is $P = \bm{d}_\Omega\bm{c}_\Omega$.

The number of iterations $q$ in the construction of the feature space $\bm{\Phi}_q$ and the dimensionality $\Omega$ of the descriptor are (hyper-)parameters of the SISSO method, to be optimized with respect to the validation error of the SISSO model, typically via a class of algorithms known collectively as {\em cross validation}, CV. See Ref. \onlinecite{SISSO}~for the CV strategy for ST-SISSO, while in section \ref{MT-SISSOcont-appl}, we discuss CV for MT-SISSO. The size of the subspace selected by SIS, $N^{\bm{S}}_{\Omega}$ is also a parameter, but not a hyperparameter to be optimized. In facts, ideally it has to be large enough to include in the set $\bm{S}_{\Omega}$ the optimal $\Omega$-dimensional solution contained in $\bm{\Phi}_q$. In practice, we invoke the relationship that the CS theory establish between size of the feature space, dimensionality of the solution, and number of data points: $N^{\bm{S}}_{\Omega} = \exp{\left(N^{\mr{M}}/(\kappa \cdot \Omega)\right)}$, where $\kappa$ is a dimensionless constant that the CS theory locates between 1 and 10. We make the further assumption that the number of features added to $\bm{S}_{\Omega}$ are the same at each iteration, i.e., $N^{\bm{S}}_{\Omega}/\Omega$. 

\subsection{Multi-task SISSO for learning continuous properties} \label{MT-SISSOcont}
We denote ${(\bm{P}^{(1)},\bm{P}^{(2)},...,\bm{P}^{N^{\mr{T}}})}$ as the set of $N^{\mr{T}}$ target property vectors, where each $\bm{P}^k$ may have a different number of samples, labelled $N^{\mr{M}}_k$.  $\bm{D}^k$ is the sensing matrix, with $N^{\mr{M}}_k$ rows and $N^{\mr{D}}$ columns, corresponding to the property $k$. Crucially, all the $\bm{D}^k$ have the same $N^{\mr{D}}$, but possibly different $N^{\mr{M}}_k$ for different properties $\bm{P}^k, k = 1, 2,..., N^{\mr{T}}$. The evaluation of the feature importance for multiple properties needs to consider the overall correlation between a feature and all the properties. 

In analogy with ST-SISSO, the MT-SISSO descriptor and model is found by the regularized minimization: 
\begin{equation}\label{eq:mtL0}
\operatorname*{arg\,min}_{\bm{C}} \sum_{k=1}^{N^{\mr{T}}}\frac{1}{N^{\mr{M}}_k}\left\Vert{\bm{P}^k-\bm{D}^k \bm{C}^k}\right\Vert^{2}_{2}+\lambda\left\Vert{\bm{C}}\right\Vert_{0},
\end{equation}
where $\bm{C}$ is the coefficient matrix, with $N^\mr{D}$ rows and $N^\mr{T}$ columns, i.e., its $k$-th column $\bm{C}^k$ is the vector of coefficients projecting $\bm{D}^k$ onto $\bm{P}^k$. The $\ell_0$ norm of the matrix $\bm{C}$ counts the number of {\em rows} that have at least one nonzero element. In practice, for each property a separate least-square regression is performed and what is minimized is the average squared error over all the regressions. The regularization imposes that when a feature $\bm{D}^*_j$ (the set of columns $j$ of all the $\bm{D}^k$) is selected (i.e., it has nonzero coefficient $C^k_{j}$) for one property $k$, then it is selected for all properties.
Mathematically, this regularization across properties (tasks) stabilizes the descriptor selection also with data unevenly distributed over the different properties. The model for any property $k$ is $\bm{P}^k = \bm{D}^k\bm{C}^k$, where each $\bm{C}^{k}$ has the nonzero elements at the same indexes $\{j_1,j_2, \ldots, j_{N^{\mr{D}}} \}$, i.e., the same features are selected for all properties. From a physical point of view, it is desirable that the different properties are homogeneous so that it {\em makes sense} that the same descriptor maps into all properties, albeit with the crucial flexibility of different fitting coefficients.

Similarly to ST-SISSO, the MT-SISSO solution of Eq. \ref{eq:mtL0} starts with a SIS step. 
To extend the SIS scheme for feature ranking with multiple properties, we first standardize all the features, i.e., the average $\overline{D^{k}_{j}}$ over all samples $N^{\mr{M}}_k$ is subtracted from each feature column vector $\bm{D}^k_{j}$ and the result is divided by its standard deviation:  $\bm{D}^k_{j} \rightarrow (\bm{D}^k_{j}-\overline{D^k_{j}})/\Vert{\bm{D}}^k_{j}-\overline{D^k_{j}}\Vert_{2}$. In this way, the absolute values of the linear correlations (scalar product) of every feature with a given property $P^k$ are comparable. We note that the standardization is the final operation {\em after} the matrices $\bm{D}^k$ are constructed following the iterative procedure described above for ST-SISSO. When the features are combined with the operators, their values are not yet standardized. \\

In the first iteration of the MT-SISSO algorithm, we have only a SIS step: the overall correlation of a feature $j$ (the $j$-th column of the sensing matrix ${D}_{k}$ for the $k$-th property) with all the properties is defined as quadratic mean of their scalar products: 
\begin{equation}\label{eq:MTSIS}
\theta_j=\sqrt{\sum_{k=1}^{N^{\mr{T}}}{<\bm{D}^k_{j},\bm{P}^k>^2/N^{\mr{T}}}}.
\end{equation}
SIS ranks the features according to $\theta_j$ and collects in $\bm{S}_{1}$ the top $N^{\bm{S}}_{1}$ features to form a subspace. Also for MT-SISSO, the feature with highest $\theta_j$ is already the optimum 1D descriptor.\\
Next, the set of residuals ${(\bm{\Delta}^{(1)}_{1},\bm{\Delta}^{(2)}_{1},...,\bm{\Delta}^{N^{\mr{T}}}_{1})}$ is evaluated, using $\bm{\Delta}^k_{1} \equiv \bm{P}^k-\bm{d}^k_{1}\bm{c}^k_{1},$, analogous to the ST-SISSO approach discussed above.

At the second and each subsequent iteration of MT-SISSO we have a SIS and a SO step. 
In the SIS step at iteration $\Omega>1$,  $\theta_j$ is evaluated as in Eq.~\ref{eq:MTSIS}, with $\bm{\Delta}^k_{(\Omega-1)}$ instead of $P^k$, and the newly selected subset of features is added to $\bm{S}_{(\Omega-1)}$ to form $\bm{S}_{\Omega}$. 

In the SO step at iteration $\Omega>1$, all possible $\Omega$-tuples in $\bm{S}_{\Omega}$ are formed.
If $\bm{d}^*_{\Omega}$ is the matrix whose columns are the members of one considered $\Omega$-tuple, $\bm{d}^k_{\Omega}$ its sub-matrix with entries related to the samples with properties $\bm{P}^k$, and $\bm{c}^k_{\Omega}=({\bm{d}^k_{\Omega}}^T\bm{d}^k_{\Omega})^{-1}{\bm{d}^k_{\Omega}}^T\bm{P}^k$ is the least-square fit of $\bm{d}^k_{\Omega}$ to  $\bm{P}^k$, then the $\Omega$-tuple that minimizes $\sqrt{(\sum_{k=1}^{N^{\mr{T}}}\frac{1}{N_k^{\mr{M}}}\Vert{\bm{P}}^k-\bm{d}^k_{\Omega} \bm{c}^{k}_{\Omega} \Vert^{2}_{2})/N^{\mr{T}}}$ is the identified $\Omega$-dimensional descriptor.

\subsection{MT-SISSO for categorical properties} \label{MT-SISSOclass}
Besides continuous properties, materials can be classified by means of categorical properties (e.g., being metal, nonmetal, topological insulator, etc.) into classes. In this work, we present MT-SISSO for classification in the following way: we consider as one {\em task} the construction of one materials-property map (with two or more classes, i.e., values of the considered categorical property). A map is a low-dimensional representation of the materials space where each material is located by means of an appropriate descriptor vector (the components of the descriptor are the coordinates in the low-dimensional representation) such that all materials sharing a certain categorical property are located in the same convex region. In a good/useful map, regions containing materials with exclusive properties (e.g., metals vs nonmetal) do not overlap. In a general materials-property map, the regions assigned to a certain class do not need to be in a convex region, actually not even in a connected region. However, in order to design a computationally efficient algorithm, we impose that the regions are convex, with some loss of generality.

The MT-SISSO formulation of the classification problem is to find multiple maps for subsets of materials that share a common descriptor, but possibly differently positioned boundaries between classes. The materials are grouped into subsets by categorical physical properties, such as bonding type, space group, etc. 
As introduced in Ref. \onlinecite{SISSO}, the mathematical formulation of ST-SISSO for classification adopts a measure of the overlap between convex regions as quantity to be minimized by the optimization algorithm. For a property with $N^\mr{C}$ classes\cite{SISSO}:
\begin{equation}\label{eq:st-l0-classification}
\operatorname*{arg\,min}_{\bm{c}} \sum_{I=1}^{N^\mr{C}-1}\sum_{J=I+1}^{N^\mr{C}}O_{IJ}(\bm{D},\bm{c})+\lambda\left\Vert{c}\right\Vert_0
\end{equation}
where $O_{IJ}(\bm{D},\bm{c})$ is the number of data in the overlap region between the $I$--domain and thse $J$--domain, $\bm{c}$ is a vector with elements 0 or 1, so that a feature $k$ (the $k$-th column of $\bm{D}$ is selected (deselected) when $c_k=1(0)$, and $\lambda$ is a parameter controlling the number of nonzero elements in $\bm{c}$. $O_{IJ}$ depends on $(\bm{D},\bm{c})$ in the sense that the nonzero values of $\bm{c}$ select features from $\bm{D}$ that determine the position (coordinates) of the data and the shape of the convex region in the map.
The MT-SISSO classification formulation for ``multi-map'' learning is simply:
\begin{equation}\label{eq:mt-l0-classification}
\operatorname*{arg\,min}_{\bm{C}} \sum_{k=1}^{N^\mr{T}}\sum_{I=1}^{N^\mr{C}-1}\sum_{J=I+1}^{N^\mr{C}}O_{IJ}(\bm{D}^k,\bm{C}^k)+\lambda\left\Vert{\bm{C}}\right\Vert_0,
\end{equation}
where a feature (a column of $\bm{D}^k$) is selected for all maps, or none, and the index $k$ runs over the tasks, i.e., the maps.

The MT-SISSO solution of Eq. \ref{eq:mt-l0-classification} involves a SIS and a SO step.
In the SIS step, the following expression is evaluated: 
\begin{equation}
\theta_j=\left(\sum_{k=1}^{N^\mr{T}}\sum_{I=1}^{N^\mr{C}-1}\sum_{J=I+1}^{N^\mr{C}}O^{1\mr{D}}_{IJ}(\bm{d}_j^k)+1\right)^{-1}
\end{equation}
where $O^{1\mr{D}}_{IJ}(\bm{d}_j^k)$ is the number of points in the overlap {\em interval} between the $I$--domain and thse $J$--domain when all data points (related to property $k$) are represented via the (one-dimensional, 1D) descriptor $\bm{d}_j^k$ (i.e., the $j$-th column of $\bm{D}^k$). In other words, all materials are projected onto a 1D coordinate, defined by each of the columns of the sensing matrix. Thinking for simplicity at only two classes $A$ and $B$, $O^{1\mr{D}}_{AB}$ counts how many points (if any) are in the overlap interval between the intervals occupied by points in class $A$ and $B$. 
The index $\theta_j$ has range (0,1], with large value corresponding to fewer data in the overlap region between domains; $\theta_j=1$ indicates no overlap between any two domains.
Similarly to the continuous-valued property case, the $N^{\bm{S}}_{1}$ features $\bm{d}_{j_1}^k, \bm{d}_{j_2}^k, \ldots, \bm{d}_{N^S_1}^k$, ,  with smallest overlap (largest $\theta_j$) are selected into the subset $\bm{S}_{1}$. Here, the ``residual'' is the set of data points in the overlap regions. This means that, at any subsequent iteration, SIS looks for the 1D feature that better classifies the data points that are not classified at the previous iterations. The newly selected features are added as usual to $\bm{S}_{(\Omega-1)}$ in order to build  $\bm{S}_{\Omega}$.

In the SO step at iteration $\Omega>1$, all the $\Omega$-tuples in $\bm{S}_{\Omega}$ are listed and the $\Omega$-tuple that minimizes $\sum_{k=1}^{N^{\mr{T}}}\sum_{I=1}^{N^{\mr{C}}-1}\sum_{J=I+1}^{N^{\mr{C}}} O_{IJ}(\bm{d}_{\Omega}^kl) $ is the selected $\Omega$-dimensional descriptor. 

Besides the domain overlap $O$, other metrics exist for classification, e.g., the number of misclassified data as defined by a support vector machine built with all the $\Omega$-tuples in $S_{\Omega}$, as adopted in Ref. \onlinecite{Bartel2018a}.

\subsection{Computational complexity of SISSO}
The time complexity for the SIS step of the SISSO algorithm is linear with the number of training data $N^\mr{M}$ and the size of feature space $N^{\mr{D}}$, i.e., $O(N^\mr{M}\! \cdot \! N^\mr{D})$,\cite{SIS}. For the SO step (in the $\ell_0$-regularization implementation as discussed in this paper), the time complexity depends on whether the target property is  continuous (regression problem) or categorical (classification problem). Though the $\ell_0$ regularization is formally NP hard, it can be made feasible by restricting to low dimension of the descriptor and moderate size of features subspace selected by SIS. With the total SIS-selected subspace size $N^{\bm{S}}_{\Omega}$ and the descriptor dimension $\Omega$, the time complexity of SO with $\ell_0$ for continuous property is $O \left( N^{\mr{M}} \cdot \left( N^{\mr{D}}\right)^2 \cdot \binom{N^{\bm{S}}_{\Omega}}{\Omega} \right)$, where $N^{\mr{M}} \cdot \left( N^{\mr{D}}\right)^2$ is the time needed for evaluating one candidate model using least-square regression and the binominal coefficient $\binom{N^{\bm{S}}_{\Omega}}{\Omega}$ is the total number of candidate models to be evaluated. For classification problems targeting two-dimensional maps, the time scaling of SO with $\ell_0$ is $O \left( \left( N^\mr{M} \right)^2 \cdot \binom{N^{\bm{S}}_{\Omega}}{\Omega} \right)$, where $\left( N^\mr{M} \right)^2$ is the time needed for evaluating one candidate model. 

\section{Results and Discussion}
\subsection{MT-SISSO for the relative stability of different structure pairs of $AB$ binary materials}\label{MT-SISSOcont-appl}
In Refs. \cite{Luca2015,Luca2017,SISSO} the learning of the relative stability between the rock-salt (RS) and zinc-blende (ZB) structures of $AB$ octet binary compounds was used as showcase study.
Here, we address, again for the octet binaries, the relative stability of 5 crystal structures, including RS and ZB and we add add 3 more crystal structures: the CsCl, NiAs, and CrB  prototypes.
The prediction of relative stability among several structures is naturally suited for MTL and in particular MT-SISSO.\\
As dataset, we use the same 82 octet binaries as in Refs. \onlinecite{Luca2015,Luca2017,SISSO}, although now each of tem was optimized the five different crystal-structure prototypes by fully relaxing all degrees of freedom compatible with the crystal symmetry (1 degree of freedom for RS, ZB, and CsCl, 2 degrees of freedom for NiAs, and 5 for CrB). Forces and energies were evaluated via density-functional theory (DFT) using the local-spin-density approximation (LSDA). The calculations were performed with FHI-aims \cite{Blum09} using the high precision third-tier basis set with ``tight settings'' for the numerical integration grids. The total energies of the data are estimated to be converged below 10 meV/atom and the energy differences between structures below 5 meV/atom. More information on these high-throughput DFT calculations can be found in Ref. \onlinecite{EmreThesis} and all inputs and outputs are in the NOMAD repository.

For the descriptor identification, we use atomic properties as input features: the ionization potential ($IP$), electron affinity ($EA$), number of valence electrons $n_\text{val}$, the group number $G$ in the periodic table, and the radii $r_{s, p, d}$ where the radial probability density of the valence $s$, $p$, and $d$ orbitals are maximal. Furthermore, equilibrium distances $d_{ij}$ of homonuclear $AA$ and $BB$, and $AB$ dimers are included. All the features were calculated with the LSDA. In the {\em NOMAD Analytics Toolkit}, also other sets of atomic features, calculated with other exchange-correlation functionals, are provided. Our experience is that the set of features used to build $\bm{\Phi}_0$ should be consistent, i.e., calculated with the same model Hamiltonian or measured with the same methodology. It is not necessarily true, however, that target properties and features in $\bm{\Phi}_0$ should be consistent. For instance, one may predict experimentally measured quantities starting from DFT features.\\
We set the parameter $\kappa$ that determines the sizes of the SIS subspaces to 3.3. 
With $N^{\mr{M}} = 82$, the subspace sizes $N^{\bm{S}}_{\Omega}$ are approximately $2 \cdot 10^5$ , $4 \cdot 10^3$, $ 5 \cdot 10^2$, and $10^2$ for $\Omega=2, 3, 4, 5$. These values are kept fixed through all our numerical test, e.g. also when $N^{\mr{M}}$ is decreased in the cross validation (CV) tests. For the routine application of ST and MT-SISSO, we note that the sizes $N^{\bm{S}}_{\Omega}$ are rahter large for the features space used in this work. We checked that even for $\kappa=4$, the same descriptors are always found at $\Omega = 2$, while for $\Omega = 3$ even  $\kappa=5$ is small enough to yield the same descriptor as for $\kappa=3.3$\\

\begin{figure}
\centering
  \includegraphics[width=.9\textwidth]{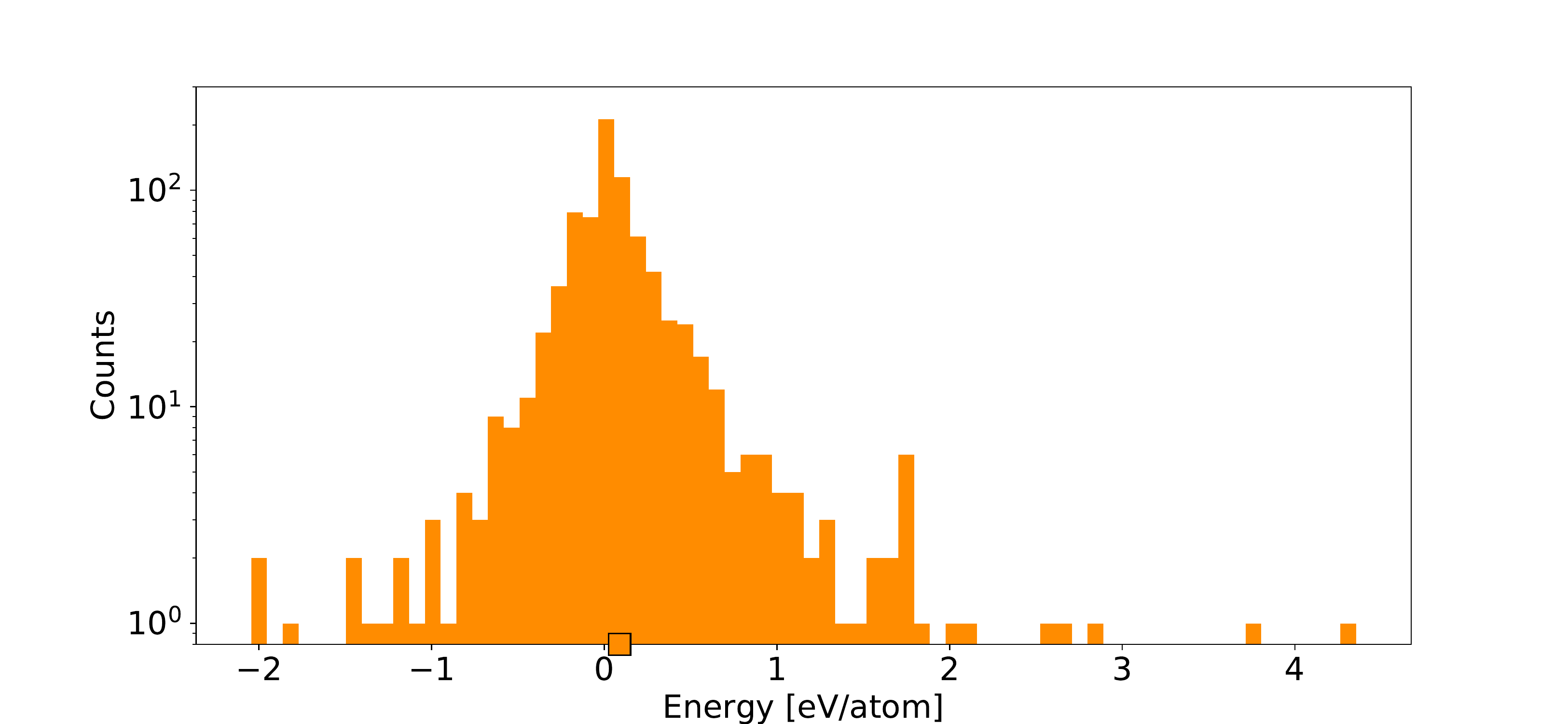}
  \caption{\footnotesize Distribution of reference (DFT-LSDA) energy differences (10 pairs of structures) for all 82 octet binaries. The square marks the average value of the distribution.}
  \label{FIG1-1a}
\end{figure}

\begin{figure}
  \centering
  \includegraphics[width=\textwidth]{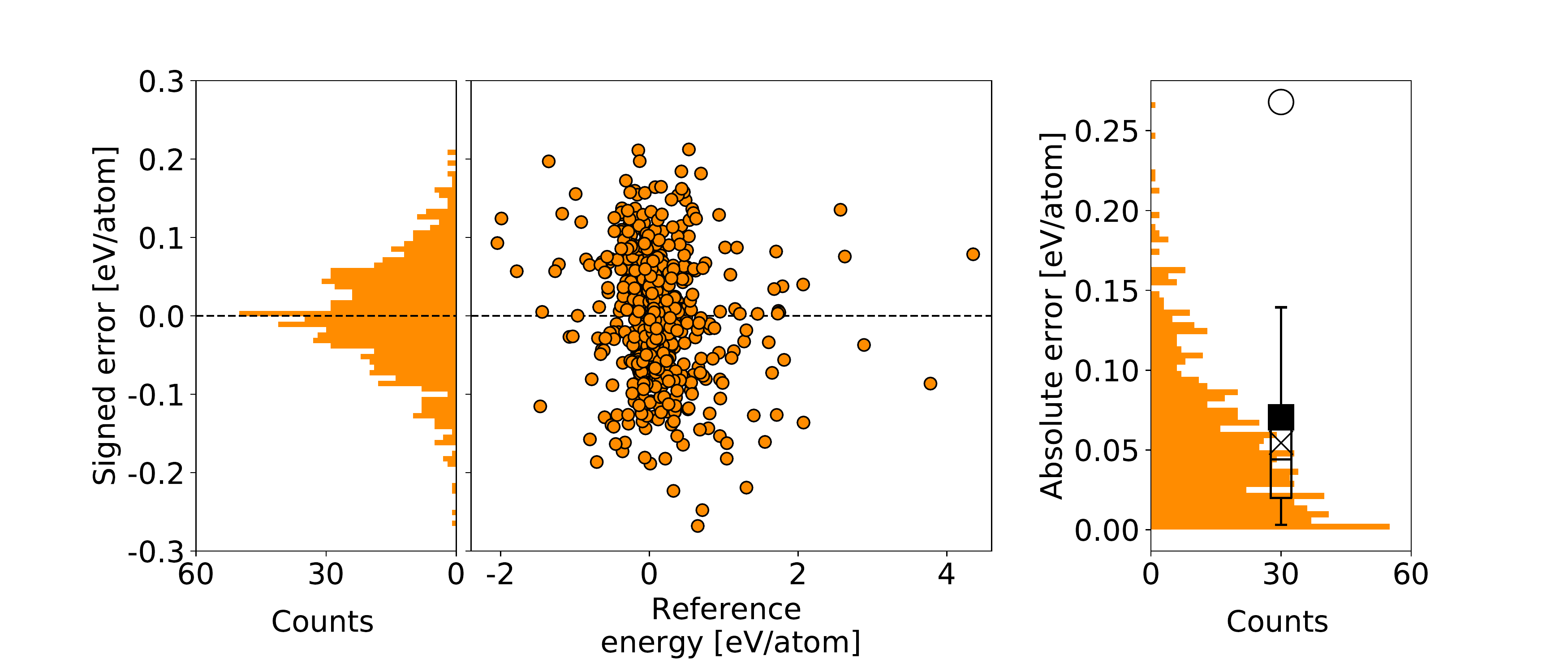}
  \caption{\footnotesize Central panel: Training errors vs. reference energies of the MT-SISSO fits to the energy differences. Left panel: distribution training errors for the same fit, obtained by integrating the central-panel plot over the reference energies. Right panel: distribution of {\em absolute} training errors and corresponding ``box plot''. The box plot marks the 25th and 75th percentiles (extrema of the rectangle), the 5th and 95th percentiles (extrema of the ``whiskers''), and the median (horizontal line inside the rectangle). Shown are also the mean absolute error (MAE, cross), the root mean square error (RMSE, solid square), and the maximum absolute error (MaxAE, circle). The feature space $\Phi_3$ and dimension $\Omega=3$ were used.}
  \label{FIG1-1b}
\end{figure}

Starting from the DFT reference cohesive energy (Total DFT energy minus the total DFT energy of the gas-phase ground-state atoms) of the five crustal structures for all the octet binary materials, we constructed 10 sets of all the possible energy differences between two crystal structures. Each energy difference is then a task in a MT-SISSO learning. 
In Fig. \ref{FIG1-1a}, we show the distribution of these energy differences. 

The main purpose of this showcase application is to learn a phase diagram (a map) where different non-overlapping regions of the diagram contain the materials with the same ground-state structure. This is similar conceptually to the classification-driven construction of materials-property maps discussed in the next section, but the crucial difference is that we target a continuous property (energy) and only {\em a posteriori} we determine the most stable phase (i.e., the ground-state crystal structure) for each material, simply by identifying which phase is predicted to have the lowest energy for each material. We emphasize that higher-energy (meta-stable) structures are learned as well. The fact that predicting energies leads to phase diagram is embedded in the fact that the MT-SISSO models are linear with the descriptor (which determines the coordinate of each material in the map), found by the MT-SISSO algorithm. With the purpose of the phase-diagram creation in mind, it should become evident why, physically, MTL is the obvious framework to use. Having one descriptor for all target properties allows to represent all the (linear) models with the same axes, resembling a traditional phase diagram with the component of the descriptor found by SISSO acting as the familiar order/control parameters.

The choice of having all the energy differences as tasks is important in order to build a phase diagram for the phase (crystal-structure) stability, when using a linear MTL like MT-SISSO. While only four energy differences (for five crystal structures) are independent, the simultaneous learning of all energy differences limits the prediction error of the relative stability between all phases. In contrast, using only one structure as reference and learning the energy difference from that structure may lead to large errors for the relative stability of any two other phases. Furtermore, a subtle implication of the MT-SISSO learning of all possible energy differences is that the models maintain an {\em internal consistency} with respect to a common energy zero. In practice, for any three structures $\alpha$, $\beta$, $\gamma$, the difference in energy $E(\alpha)-E(\gamma)$ is by construction equal to $(E(\alpha)-E(\beta))-(E(\gamma)-E(\beta))$. This is not (necessarily) true if the three energy differences are learned with separate, independent models. We will come back to this aspect when discussing the phase diagram derived from the learned MT-SISSO models.

In Fig. \ref{FIG1-1b}, we show the training errors of the MT-SISSO model for the energy differences, trained by using the feature space $\bm{\Phi}_3$ and dimensionality $\Omega=3$ (see further for the justification of this choice). The overall RMSE errors, 0.07 eV/atom, should be compared to the standard deviation of the reference-data distribution, which is 0.49 eV/atom. The latter value represents the so-called {\em baseline}, i.e., the RMSE for the model that predicts for all points the average values of the target property over the training data. 

Here, we note that the MT-SISSO approach can be also seen as a way to include collective or structural features of the materials, such as the local environment of each atom, in the learning scheme. Rather than trying to explicitly include a functional dependence of the local environments, the different environments (here, the different crystal-structure prototypes) are assigned to different tasks and each to each local environment is assigned a different set of coefficients for the mapping of the common (environment-independent) descriptor found by MT-SISSO to the different tasks.

\begin{figure}
\centering
  \includegraphics[width=.8\textwidth]{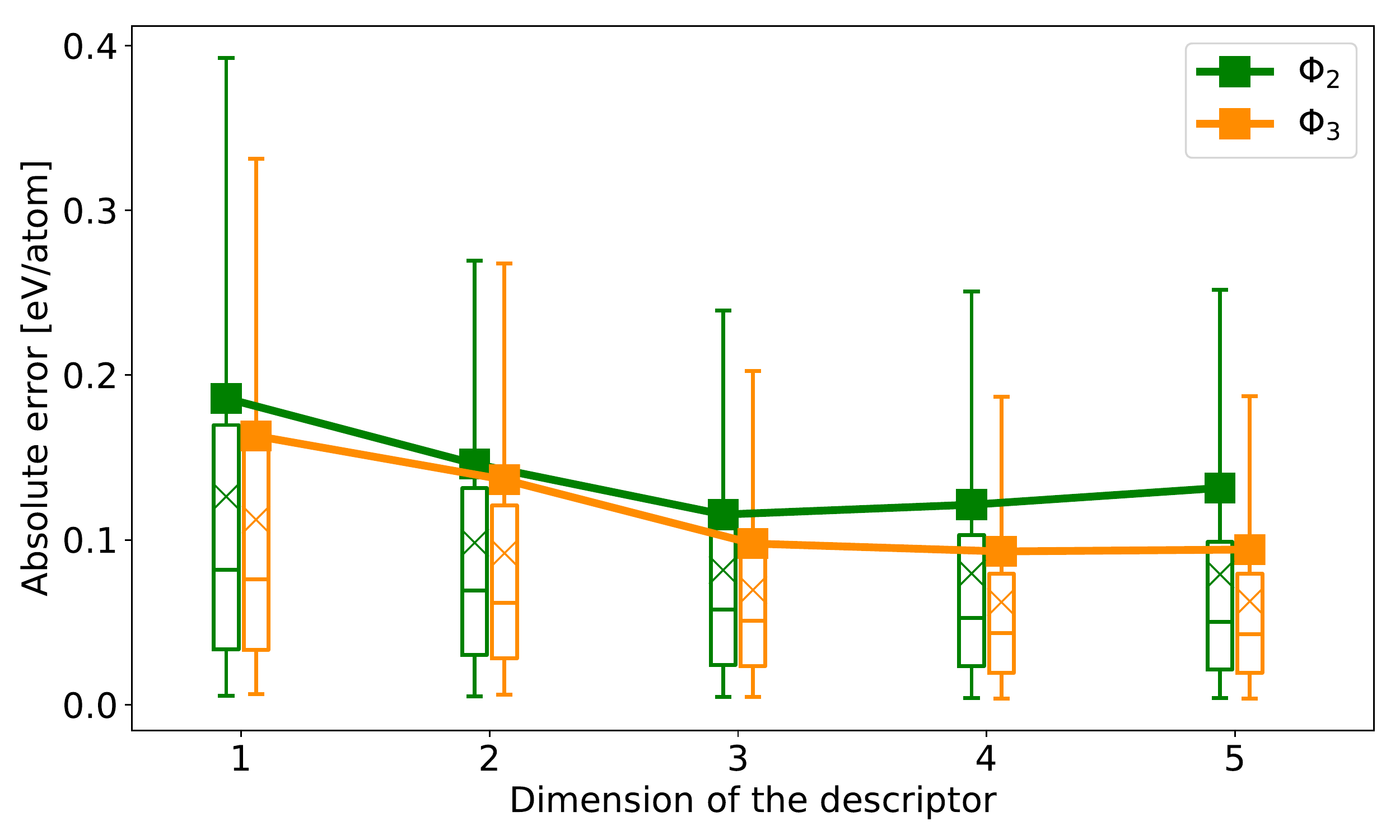}
 \caption{\footnotesize Cross-validation prediction errors of MT-SISSO models for the energy-difference learning, as function of the dimension of the descriptor $\Omega$, for the feature space $\Phi_2$ and $\Phi_3$. All errors are averaged over 30 repetitions of leave-10\%-out CV (MT-SISSO is trained over 90\% of randomly selected data and tested on the remaining 10\%). The box plots and symbols are consistent with Fig. \ref{FIG1-1b}, right panel.}
 \label{FIG1-2a}
\end{figure}

In Fig. \ref{FIG1-2a} (the corresponding numerical values are tabulated in Table \ref{TB1-1}), we show the cross-validation (CV) test for the energy-difference learning, performed in order to assess the two hyperparameters of MT-SISSO: the (size of the) feature space $\bm{\Phi}_q$ and the dimensionality $\Omega$ of the descriptor. To the purpose, we performed a leave-10\%-out CV, i.e., 10\% of the materials are left out of the training set, the MT-SISSO model is trained on the remaining 90\% of the materials, and the errors are measured for the left-out materials. This random selection of training and validation sets was repeated 30 times, which we found sufficient to converge the validation RMSE to 0.01 eV. We note a) that all the 10 target properties of a material are excluded from the training set when it is left out and b) the standardization of the features is performed at each random selection of the training set, only on the features relative to the actual training data points. This latter highly recommended practice is crucial to avoid information ``contamination'' between the training and validation set.

\begin{figure}
\centering
  \includegraphics[width=1.\textwidth]{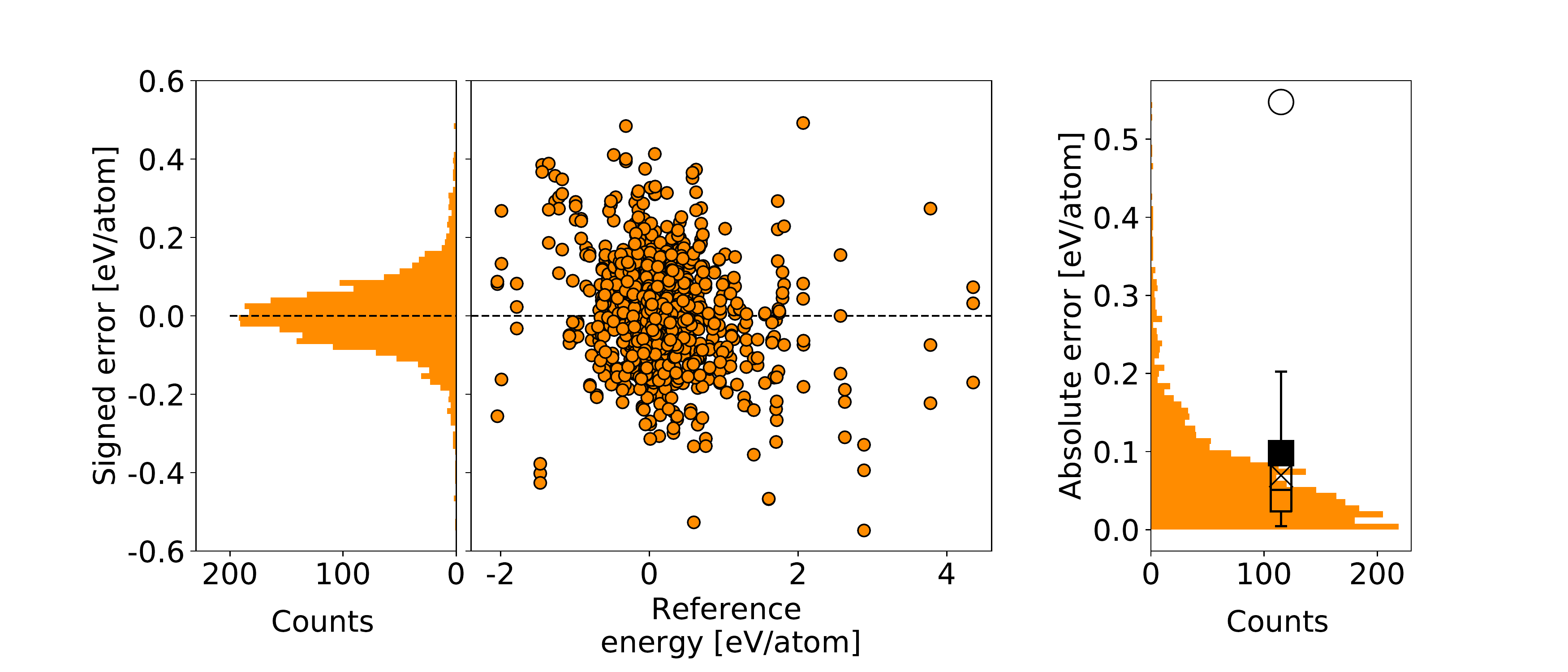}
  \caption{\footnotesize Same set of quantities as in Fig. \ref{FIG1-1b}, for the signed and absolute {\em prediction} errors for ($\Phi_3,\Omega=3$).}
  \label{FIG1-2b}
\label{FIG1-2}
\end{figure}

Analysis of Fig. \ref{FIG1-2a} reveals that models trained by using the larger feature space $\bm{\Phi}_3$ (containing $\sim 2 \cdot 10^{10}$ features) are consistently better performing (in terms of prediction errors) than models trained starting form $\bm{\Phi}_2$ (containing $\sim 2.4 \cdot 10^{5}$ features), for all dimensions. Root mean square errors (RMSE) and mean absolute errors (MAE) are only marginally better when going from $\bm{\Phi}_2$ to $\bm{\Phi}_3$, but we notice that the largest percentiles (75th and 95th) improve significantly, especially for $3 \leq \Omega \leq 5$. Looking at larger percentiles of the error distributions, besides looking at mean errors, is important because, for a predictive model, we are typically interested that the worst cases still yield relatively small errors. 
The overall best model is $(\bm{\Phi}_3,\Omega = 5)$, but we also notice that, for $\bm{\Phi}_3$, the improvement of all error indicators when going from $\Omega=3$ to $\Omega=5$ is only marginal. Therefore, in view of the significantly smaller computational time needed to train $\Omega=3$ vs $\Omega=5$, in the following tests, we focus on $(\bm{\Phi}_3,\Omega =3)$, starting from Fig. \ref{FIG1-2b}, where we report the detailed analysis of the signed and absolute errors for these latter settings.

\begin{table}
\begin{center}

  \begin{tabular}{ l | c || c c  c  c  c} 
                          & $\Omega$ & RMSE & Median & $p_{75}$ & $p_{95}$ & MaxAE \\ \hline \hline
\multirow{5}{*}{$\bm{\Phi}_2$} & 1 & 0.186 & 0.082 & 0.170 & 0.393 & 1.098 \\
                          & 2 & 0.146 & 0.069 & 0.131 & 0.272 & 1.055 \\
                          & 3 & 0.115 & 0.058 & 0.112 & 0.240 & 0.649 \\
                          & 4 & 0.121 & 0.053 & 0.103 & 0.252 & 0.968 \\
                          & 5 & 0.132 & 0.050 & 0.099 & 0.252 & 1.385 \\ \hline
\multirow{5}{*}{$\bm{\Phi}_3$} & 1 & 0.163 & 0.076 & 0.158 & 0.332 & 1.056 \\
                          & 2 & 0.137 & 0.062 & 0.121 & 0.268 & 0.973 \\
                          & 3 & 0.098 & 0.051 & 0.090 & 0.205 & 0.548 \\
                          & 4 & 0.093 & 0.043 & 0.079 & 0.187 & 0.742 \\
                          & 5 & 0.094 & 0.043 & 0.080 & 0.189 & 0.709 \\

\end{tabular} 
\end{center} 
 \captionof{table}{\footnotesize Tabulated values from Figure~\ref{FIG1-2a}. $p_{75}$ and $p_{95}$ are the 75th and 95th percentiles, respectively, RMSE is the root mean square error, and MaxAE is the maximum absolute error. All quantities are given in [eV/atom].}
\label{TB1-1}
\end{table}

\begin{figure}
\includegraphics[width=0.9\textwidth]{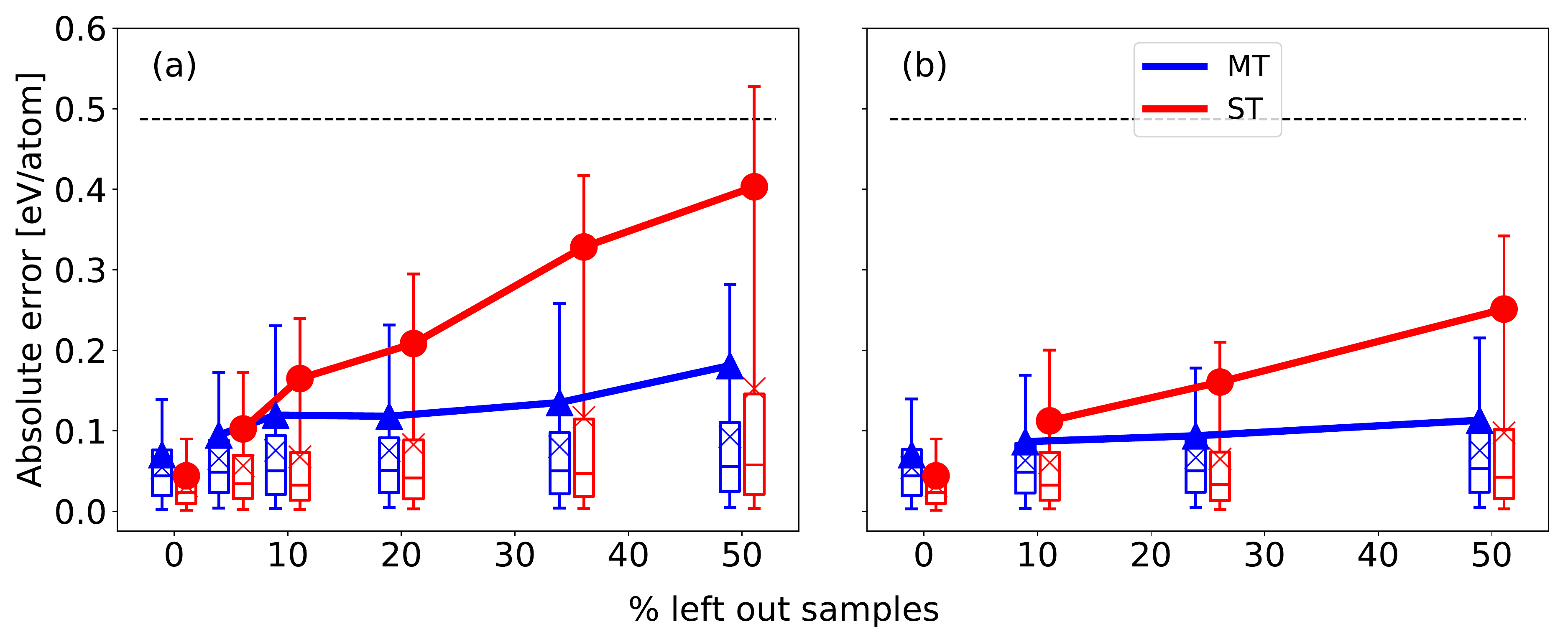}
\caption{\footnotesize Prediction errors of MT-SISSO vs (average) ST-SISSO for (left panel) ``leave $x$\% of (materials, structure) data out'' and (right panel) ``leave $y$\% of data for one crystal structure out''.  The symbol convention is the same as in Fig. \ref{FIG1-1b}. In both panels, the errors at 0\% data out in both panels are training errors. The horizontal line at 0.49 eV is the baseline (see text).
\label{FIG1-3}}
\end{figure}

We now turn our attention to two tests that reveal the peculiarity of MTL vs traditional ST learning when only incomplete data are available. 
In the first test, we selected left-out sets in this way: one material and one crystal structure are randomly selected and the all the energy differences involving the selected  structure are eliminated from the training set for the selected material. The procedure is repeated until a prefixed $x$\% of pairs  (material, structure) are eliminated (we recall the total number of such pairs is $82\times5=410$). This test simulates the training over a materials database where for some (or many) materials the information for only some crystal structures is available. It would be of great value if from such dishomogenous database, one could predict the missing information. For a meaningful test, we added the following two constraints in the simulated elimination of database fields: for each material, the energy of at least 2 crystal structures is known and for each of the 10 tasks (energy differences) there are at least 4 materials carrying the information, in order to have enough data to train the 4 fitting coefficients of the $\Omega=3$ model. For each $x$\% selected value, we train one MT-SISSO model and 10 independent ST-SISSO models (one for each task of MT-SISSO). We then look at the prediction errors on the missing data.
Figure~\ref{FIG1-3}a shows the outcome of the test. With abuse of notation, the values at 0\% refer to training error. As one should expect, ST-SISSO yields lower training error due to higher flexibility (for each task, a different descriptor can be chosen). However, as soon as data are missing, MT-SISSO rules with lower RMSE and, crucially, with lower largest errors. Interestingly, the quality of MT-SISSO stays pretty unchanged, for all error indicators, over a wide range of amount of missing data. 

In the second test, we selected one crystal structure (here, RS) and then we removed the energy values for a given $y$\% of materials. Removing the energy value of one structure implies the removal of 4 energy differences from the (material, energy differences) database. One MT-SISSO model and 4 ST-SISSO models are trained and the errors for the selected structures are evaluated on the missing materials This test simulates the case of a new crystal structure being identified for only few materials in the database and one wants to learn with the fewest possible data the predicted energy in such new crystal structure for all materials. Figure~\ref{FIG1-3}b shows the performance of the MT-SISSO model vs the average of the 4 ST-SISSO model. Again the training error (at 0\%) favors ST-SISSO and again MT-SISSO's performance remain impressively constant over a wide range of amount of missing data.

These two tests show numerically what should be expected from a physical point of view: It is reasonable to assume that the energy of different crystal structures depend on the same mechanism encoded in the properties of the gas-phase atoms used as primary features. Therefore MT-SISSO uses at best the (possibly scarce) information scattered over all crystal structures to identify such mechanism. In this way the prediction on the scarcely known materials and/or crystal structures is more reliable than a model that uses information from only one crystal structure (or, one pair of crystal structures, as in the presented case) to identify the descriptor.

\begin{figure}
\begin{subfigure}{.5\textwidth}
  \includegraphics[width=\linewidth]{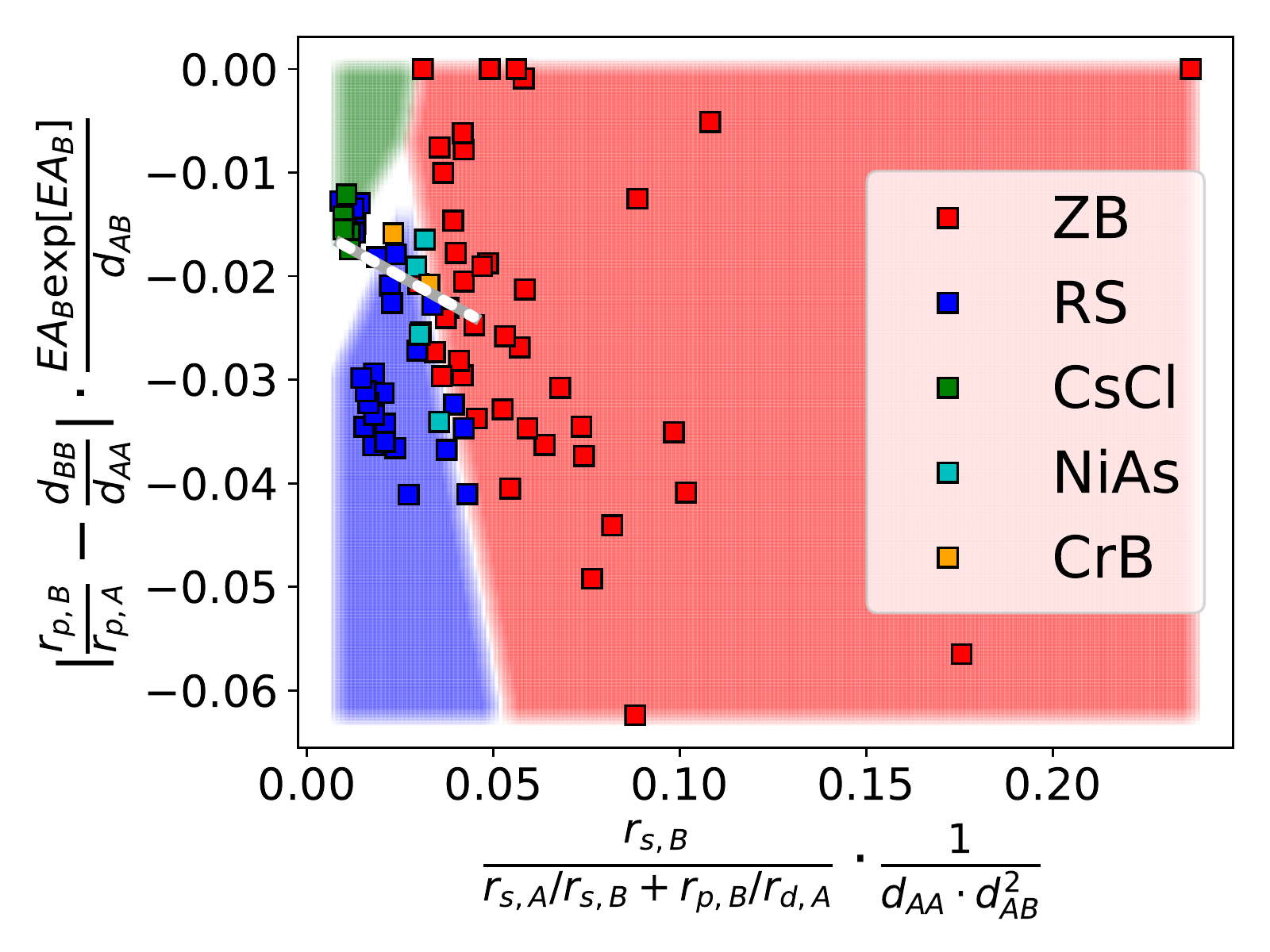}
  \label{FIG1-4a}
\end{subfigure}%
\begin{subfigure}{.5\textwidth}
  \includegraphics[width=\linewidth]{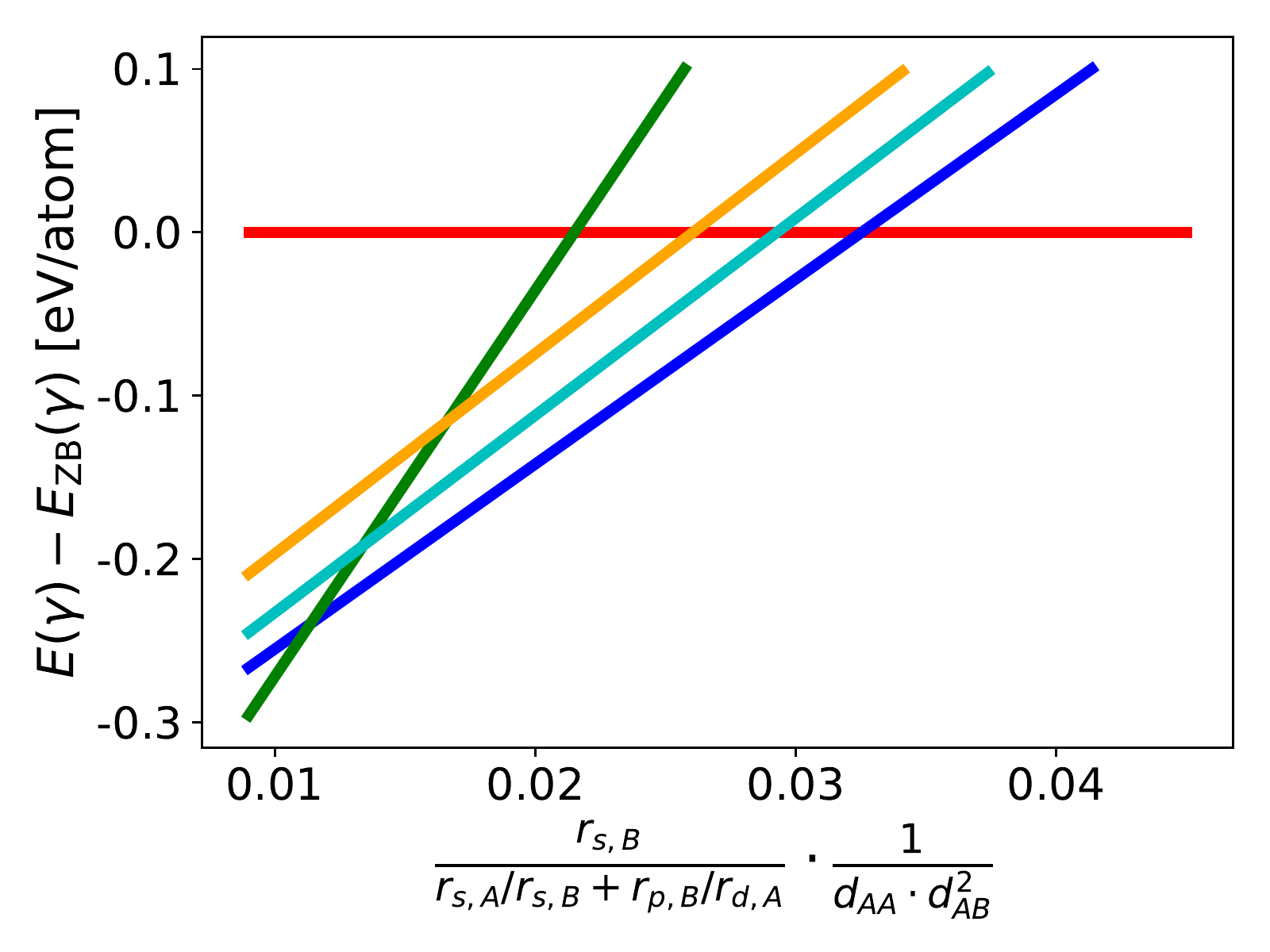}
  \label{FIG1-4b}
\end{subfigure}
\caption{\footnotesize Left: MT-SISSO-learned phase diagram (structure map) of the ground-state crystal structure for the octet binaries. The colored areas represent the predicted stability region for the structure with the same color in the legend. The squares are colored according to the reference lowest-energy structure. The white color marks areas where the difference between the energy of the lowest-energy and the second lowest-energy structures differ by at most 0.03 ev/atom. Right: cut of the phase-diagram along the dashed line shown in the left panel. The lines are the traces of the planes representing the predicted energy difference from the baseline (ZB structure).}
\label{FIG1-4}
\end{figure}

We close the section on MT-SISSO by showing how the ($\Omega=2$) MT-SISSO model trained over all data points can be used to draw a phase-diagram (crystal-structure map). The model identified by MT-SISSO for each task can be represented as a plane in a 3D space, where the coordinates $(x,y)$ are the components of the descriptor and coordinate $z$ is the predicted energy. The mentioned property of {\em internal consistency} among MT-SISSO models for (energy) differences allows for the unambiguous determination of the predicted lowest-energy structure for each coordinate $(x,y)$. A color is associated with any specific crystal structure and assigned to a square (pixel) $(\delta x,\delta y)$ centered on $(x,y)$ when the corresponding structure is the lowest in energy at $(x,y)$.
Figure~\ref{FIG1-4}a represents the structure map for the octet binaries. The colored area refer to the predictions and the colored squares are the reference data. The white color marks areas where the energy difference between the lowest-energy and the second lowest-energy structures differs by less than 0.03 eV/atom. In order to give an insight into the 3D visualization of the structure map, we show in Fig. \ref{FIG1-4}b, a cut along the gray-white dotted line marked in Fig.~\ref{FIG1-4}a. This show that some crystal structures are predicted to be very close in energy for certain values of the descriptors. In a realistic application, one may conclude that the actual ground state in the neighborhood of those values of the descriptor may be any of the low-energy structures (in particular, at finite temperature), while those that are predicted to be very high in energy can be safely discarded as candidate ground state. To gauge the trustfulness of the presented phase diagram, we mention that the largest prediction error for a structure that appears ``misclassified'' (the color of its symbol does not match the background --- predicted --- color) is 0.09 eV/atom.

\subsection{MT-SISSO for the metal/insulator classification of $A_xB_y$ binary materials}\label{MT-SISSOclass-appl}
\begin{figure}
\includegraphics[width=0.6\textwidth]{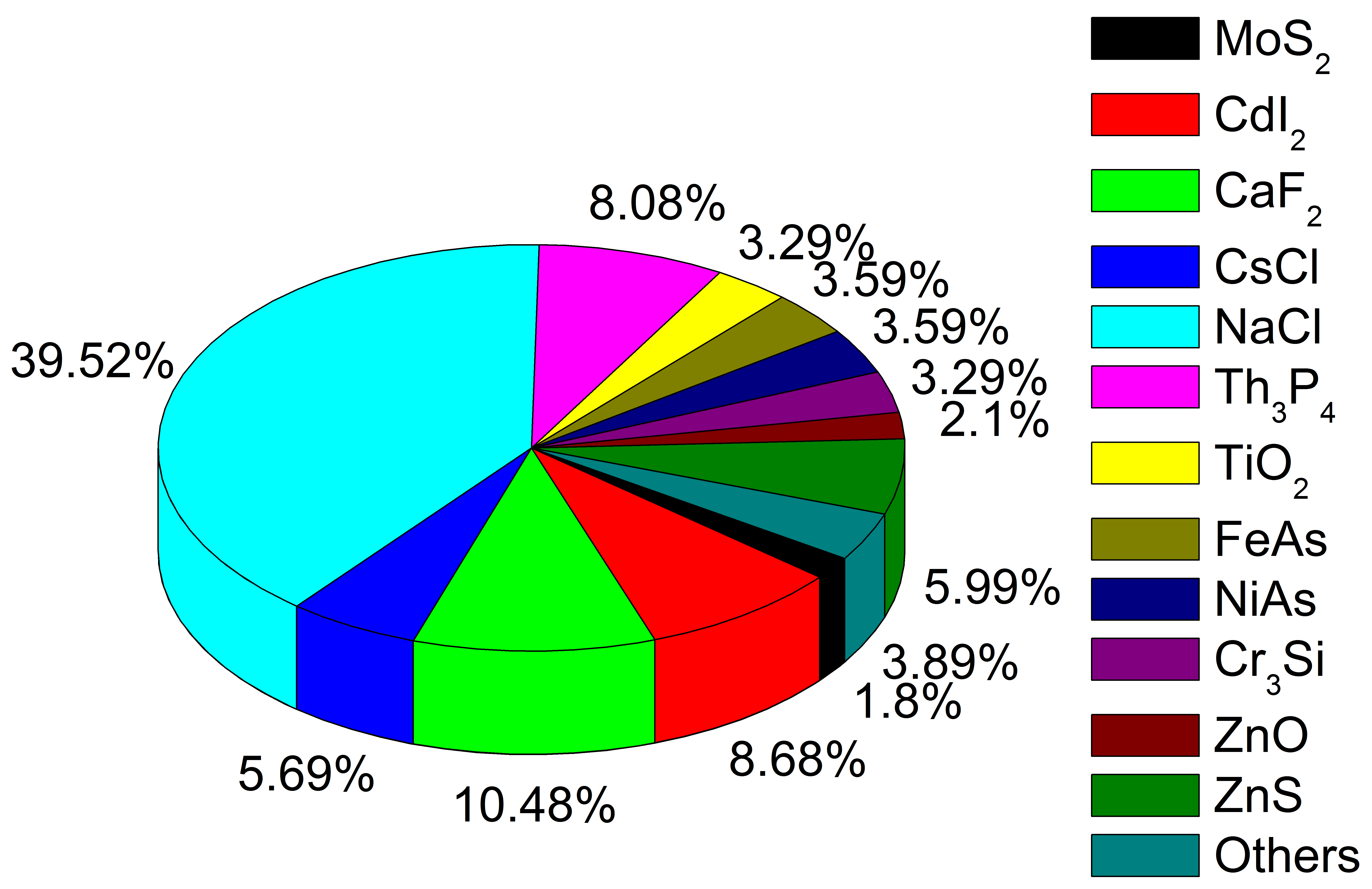}
\caption{\footnotesize Pie chart showing the distribution of the 334 reference binary materials, taken from SpringerMaterials, over the 17 considered crystal-structure prototypes.\label{FIG2-0}}
\end{figure}

In Ref. \onlinecite{SISSO}, a SISSO-trained model for the metal/insulator classification of 299 binary materials distributed over 15 prototypes was presented, with (experimental) reference data collected from the SpringerMaterials database\cite{springermaterials}. That model achieved 99\% classification accuracy with a 2D descriptor, but had several constraints, i.e., ignoring materials of certain bonding types. In the present work, we extend the metal/insulator dataset to totally 334 $A_xB_y$ binary materials (197 metals and 137 nonmetals) belonging to 17 crystal-structure prototypes. The new dataset includes the 15 three-dimensional prototypes previously considered\cite{SISSO} and, in addition, two layered prototypes: CdI$_2$ and MoS$_2$.
The pie-chart of the distribution of data points over prototypes is shown in Figure~\ref{FIG2-0}. The descriptor described in Ref. \onlinecite{SISSO} was a function of properties of gas-phase atoms plus one collective feature, namely the unit-cell volume. At first, by using the same set of primary features, we check whether SISSO can find a single map that correctly classifies into metal vs nonmetals the materials in all 17 prototypes. Specifically, we considered as primary features: \{ionization energy $IE$, Pauling electronegativity $\chi$, covalent radius $r_{cov}$, unit cell volume normalized by total atom volume $V_{\rm cell}/\sum{V_{\rm atom}}$, bonding distance in the material between $A$ and $B$ $d_{AB}$, coordination number of $A$ species $N^{\mr{N}}_{A}$ and of $B$ species $N^{\mr{N}}_{B}$, and atomic fraction for $A$ $x_A$ and $B$ $x_B$\}. As in Ref. \onlinecite{SISSO}, the values for the atomic features are taken from WebElements \cite{webelements} and the information for building the structural features (atomic coordinates, species, and lattice vectors) comes from the SpringerMaterials\cite{springermaterials} database.  Furthermore, we considered as operator set: $\{+,-,\times,/,\exp{},\log{},\vert-\vert,\sqrt{\phantom{0}},^{-1},^{2},^{3}\}$. From these ingredients, we build the feature sapce $\bm{\Phi}_3$. The size of the SIS-selected subspace for each descriptor dimension was set to 10$^4$ which is a big yet manageable size for descriptors up to 2D. Unless otherwise stated, these settings are used for all the classification problems discussed below.

\begin{figure}
\includegraphics[width=0.6\textwidth]{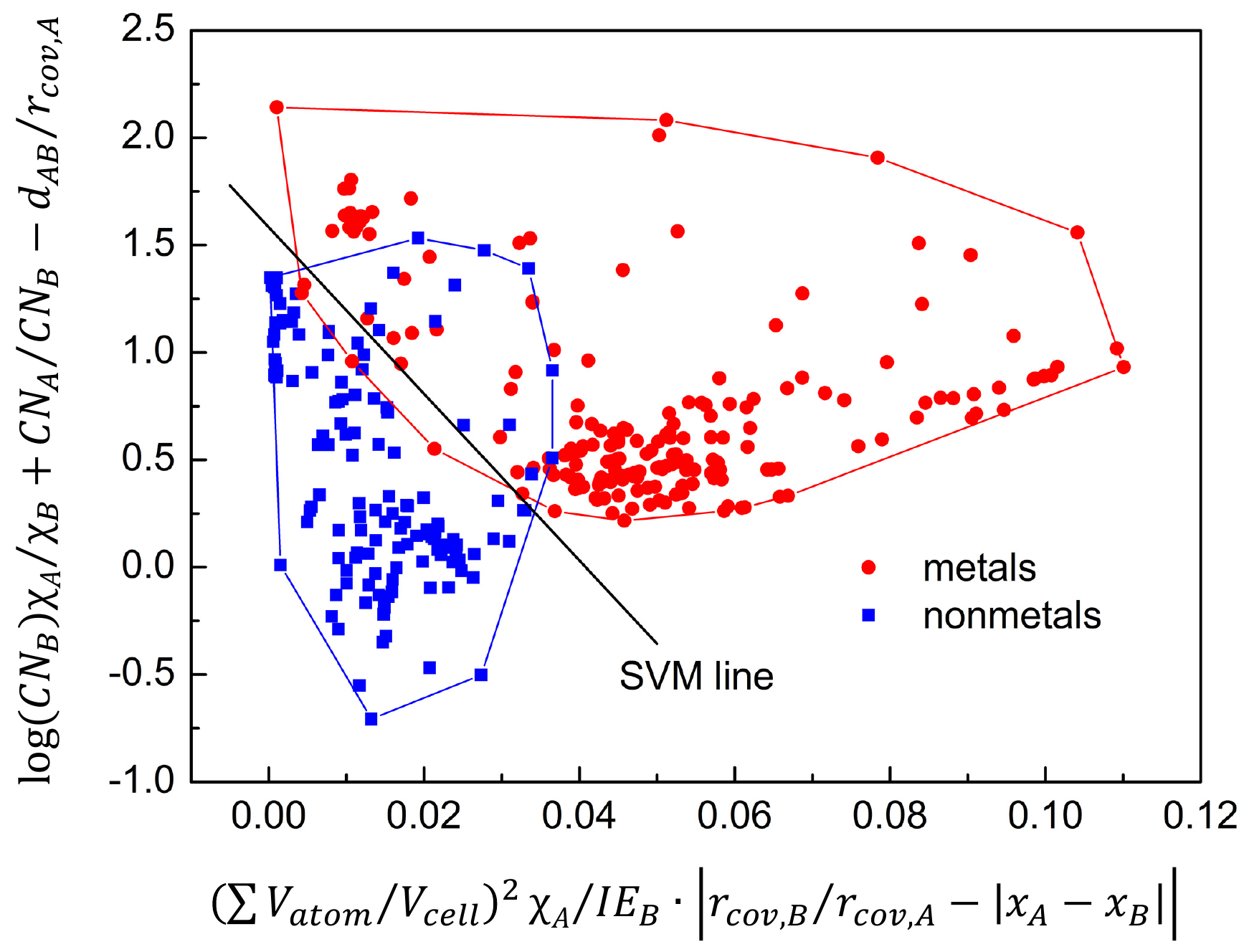}
\caption{\footnotesize The metal/nonmetal classification map for binaries on all 17 prototypes. The (red-) blue-bordered convex regions denote the (metal) nonmetal domain. The linear-SVM-trained separation line was found with the 2D descriptor fixed to the one found by SISSO. 
\label{FIG2-1}}
\end{figure}

Figure~\ref{FIG2-1} shows the classification map by the best SISSO-trained 2D descriptor. 
There is an overlap between the metal and nonmetal regions, and in total there are 36 data points in the overlap region. Among the materials in the overlap, 13 (8 metals and 5 nonmetals) are in the CdI$_2$ prototype, and 6 (1 metal and 5 nonmetals) are in the MoS$_2$ prototype. For the latter prototype, we have information only on 6 materials. The other 17 materials in the overlap belong to the other 15 prototypes. 
In the map of Figure~\ref{FIG2-1}, the optimal separation line was found by using a linear support vector machine (SVM) with the SISSO-determined 2D descriptor. According to the SVM metric, 17 out of 334 materials are misclassified. To avoid confusion, in the following the number of misclassified data points will always refer to the SVM metric, while as SISSO figure of merit we report the ``number of data point in the overlap region''. It is not strictly necessary to apply SVM after SISSO, as SISSO for classification already targets a map that separates as much as possible (ideally, fully, without overlap) the different classes of materials. However, the SISSO model is determined by all the boundary materials defining the convex regions. An SVM line (at fixed descriptor determined by SISSO) is a well defined and a much simpler model, which does not conflict with the SISSO model.

Though a global descriptor (up to 2D) for the accurate metal/insulator classification of all prototypes is not found with the current primary features, the independent classification for each prototype with 100\% training accuracy is very easy to achieve. Table~\ref{TB2-1} shows the simple 1D descriptors for 100\% classification of metal/insulator of the binary materials for each prototype independently. Actually, ST-SISSO finds many descriptors for the 100\% classification within each prototype, and Table~\ref{TB2-1} shows only the most simple ones (with least number of mathematical operators in the features). However, we note that many prototypes have very few data points and therefore the classification model risks to be overfit. 

\begin{table*}
\caption{\footnotesize Descriptors yielding metal/nonmetal 100\% classification accuracy within each prototype. The primary features of $IE$, $\chi$, $V_{\rm cell}/\sum{V_{\rm atom}}$, and $d_{AB}$ were used for these calculations; coordination number $N^{\mr{N}}$ and atomic fraction $x$ were excluded because they are constant within one prototype. Since all descriptors are one dimensional, we also provide the threshold values for the metal/nonmetal transition (metals are for values of the descriptor smaller than the threshold).\label{TB2-1}}
\begin{tabularx}{\textwidth}{X|c|c|c}
\hline 
prototype\footnote{ReO$_3$ prototype was not considered because of only one metal and one nonmetal available.} & number of data & descriptor & boundary \\ \hline \hline
MoS$_2$ & 6 (1 metal, 5nonmetals) & $\chi_A$  & 1.68\\ \hline
CdI$_2$ & 29 (8 metals, 21 nonmetals) & $d_{AB}\chi_B^3$  & 41.08\\ \hline
CaF$_2$ & 35 (21 metals, 14 nonetals) & $\chi_B$  & 2.68\\ \hline
CsCl    & 19 (16 metals, 3 nonmetals) & $IE_B$ & 9.55 \\ \hline
NaCl    & 132 (87 metals, 45 nonmetals) & $\frac{V_{\rm cell}}{\sum{V_{\rm atom}}}\frac{IE_{A}IE_Br_{covA}}{\chi_A}$  &135.79 \\ \hline
Th$_3$P$_4$ & 27 (23 metals, 4 nonmetals) & $\frac{V_{\rm cell}}{\sum{V_{\rm atom}}}IE_{A}(d_{AB}IE_B)^2$  & 676.24 \\  \hline
TiO$_2$  &  11 (2 metals, 9 nonmetals) & $-\chi_A $ & -2.105 \\ \hline
\{FeAs,NiAs,ThH2,Cr$_3$Si,ZnO, 
ZnS,Al$_2$O$_3$,La$_2$O$_3$,SiC \}\footnote{The prototypes that has either only metals or only nonmetals were grouped as a mixed ``prototype".} & 73 (38 metals, 35 nonmetals) & $\frac{V_{\rm cell}}{\sum{V_{\rm atom}}}IE_B\chi_B$  & 42.90\\ \hline
\end{tabularx}
\end{table*}

\begin{figure}
\includegraphics[width=\textwidth]{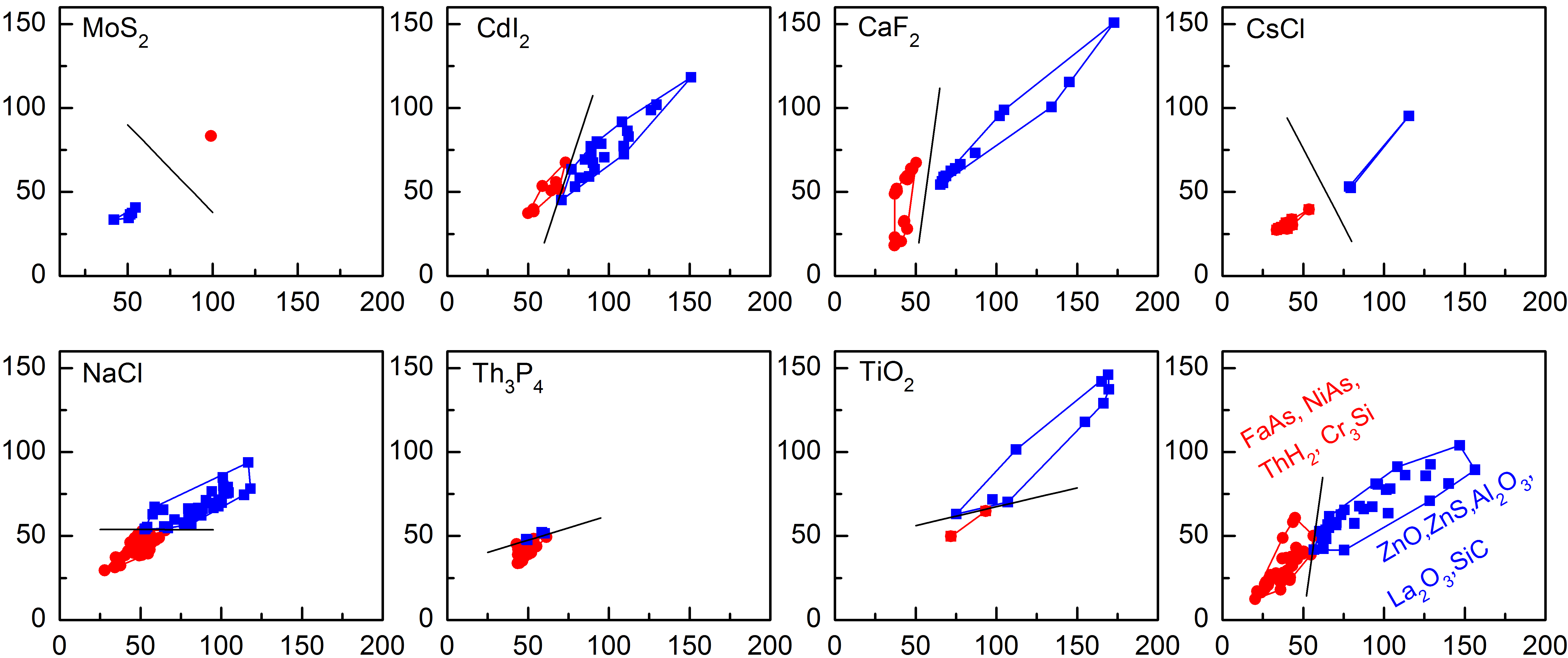}
\caption{\footnotesize MT-SISSO results for the classification of metal/nonmetal for 17 crystal prototypes. The component of the SISSO-determined descriptor on the $x-$ and $y-$axes (the same in all plots) are given in Eq. \ref{eq:mtdescexp}. There is zero overlap between the metal and nonmetal domains on all the maps. The separation lines were found via linear SVM.}
\label{FIG2-2}
\end{figure}

MT-SISSO mediates between the two extrema of the global, inaccurate map and the one-per-prototype map, that is probably overfit for prototypes for which few data points are available. Interpreting the map for one prototype as one task, MT-SISSO can be set up to look for a set of maps, all  defined by the same descriptor, but with differently located convex regions for the classification.
We ran MT-SISSO for classification with the same parameter settings as for the global descriptor, except that the prototype ReO$_3$ is excluded (this prototype is represented by only 1 metal and 1 nonmetal in our reference dataset) and the crystal features $x_A$, $x_B$, $N^{\mr{N}}_A$, and $N^{\mr{N}}_B$ are removed because they are constant within a given prototype. Figure~\ref{FIG2-2} shows the MT-SISSO maps. Overall and individually, they achieve perfect classification. The common 2D descriptor is:
\begin{align}
\label{eq:mtdescexp}
\begin{split}
d_1 = & \frac{V_{\rm cell}}{\sum{V_{\rm atom}}}\frac{\chi_{B}\exp{(r_{cov,A})}}{\chi_{A}r_{cov,A}}   \\
d_2 = & \frac{V_{\rm cell}}{\sum{V_{\rm atom}}}IE_{A}IE_{B}r_{cov,A}\sqrt{\chi_{A}/\exp{(\chi_{A})}} 
\end{split}
\end{align}
We note that this descriptor has similar ``ingredients'' (primary features) as the global ST-SISSO descriptor presented in Ref. \cite{SISSO}, in particular the descriptor depends linearly on the inverse of the packing fraction ${\sum{V_{\rm atom}}}/{V_{\rm cell}}$, which is the only selected collective feature, i.e., related to the actual atomic structure of the material. 

\begin{figure}
\includegraphics[width=0.8\textwidth]{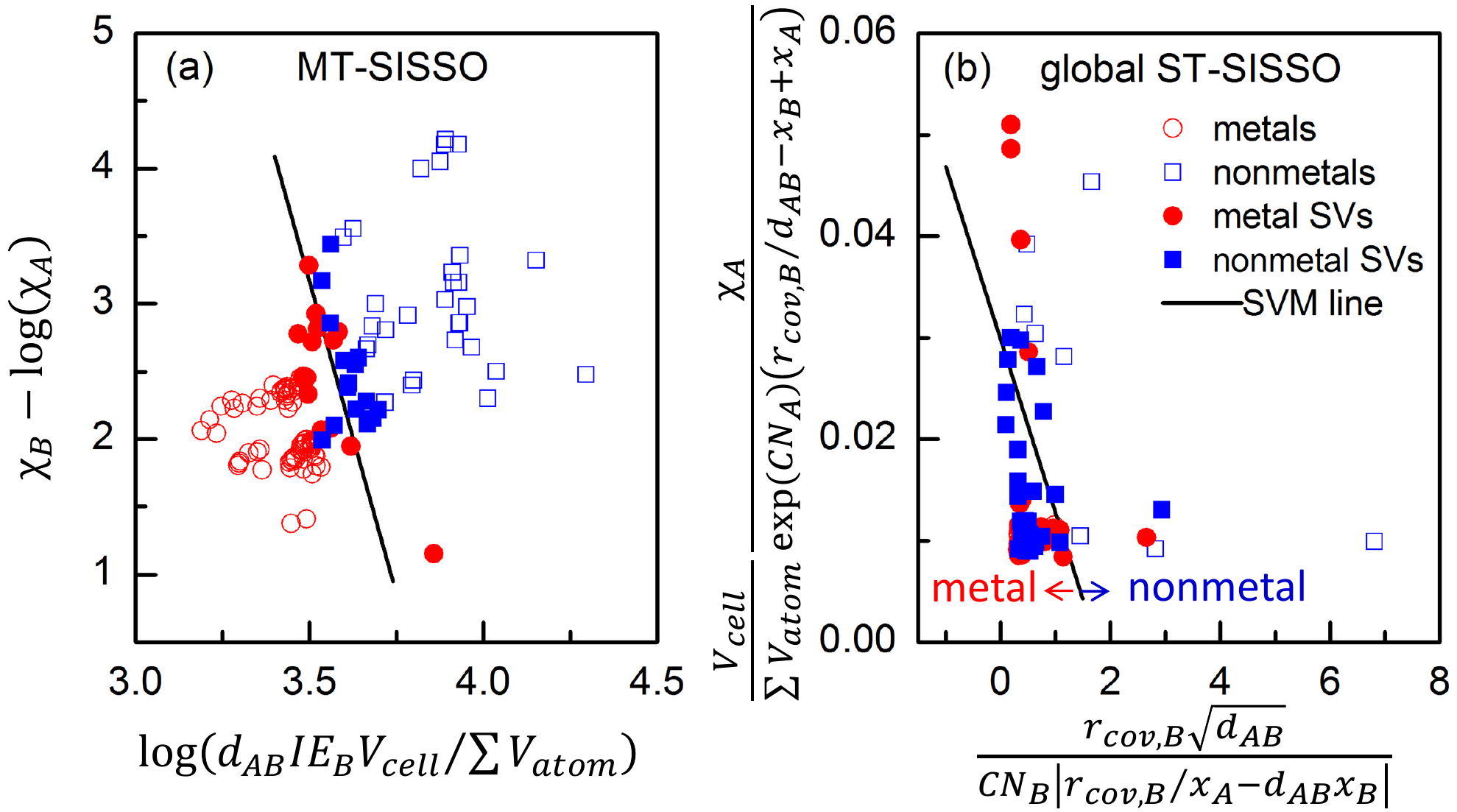}
\caption{\footnotesize ``Leave-rock-salt-prototype-out'' generalization test for MT-SISSO (left panel) and global ST-SISSO (right panel). Both MT-SISSO and ST-SISSO descriptors are trained on all prototypes except rock-salt and the support-vector-machine (SVM) line is trained on the rock-salt data, at fixed descriptor. The filled symbols are the support vectors (SV) of the SVM model. 
\label{FIG2-3}}
\end{figure}

To demonstrate the generalizability of MT-SISSO descriptors on unseen prototype materials, we performed a ``leave-one-prototype-out'' validation. In practice, we focused on the RS prototype (that includes about 40\% of the training dataset) and we trained the metal/nonmetal classification wih MT-SISSO and with global ST-SISSO. The latter is ST-SISSO by using all training data to train a single metal/nonmetal map. This is the same approach as in Ref. \cite{SISSO}, where however fewer prototypes were considered. For ST-SISSO, the features coordination number $N^{\mr{N}}$ and atomic fraction $x$ are included as primary features in $\bm{\Phi}_0$. Subsequently the RS data are projected into the 2D descriptor determined by the training on the other prototypes and a SVM model is trained at fixed descriptor. We name these two approaches MT-SISSO+SVM and ST-SISSO+SVM. In this test, we have omitted the ST-SISSO learning on one prototype because all the data points of the left-out prototype are left out of training at the SISSO stage. 
The results are shown Fig. \ref{FIG2-3}. The descriptor identified by global ST-SISSO scatters metals and nonmetals NaCl binaries all around the map, making a classification impossible. In contrast, the MT-SISSO descriptor yields a map that separate fairly metals vs nonmetals, without having access to any direct information on RS materials in the training. Quantitatively, the number of misclassified NaCl materials by MT-SISSO+SVM is 6 out of 132 and one can appreciate by naked eye in Fig. \ref{FIG2-3}a that the misclassification is not ``severe'', i.e., the misclassified materials are close to the SVM line. For ST-SISSO+SVM the number of misclassified materials is 36 out of 132 and visual inspection (Fig. \ref{FIG2-3}b) reveals that, without the labels ``metal'' (``nonmetal'') in the half planes, it would be even difficult to decide which side of the line is predicted to contain metals (nonmetals). \\
We repeated the test for other prototypes, but, mainly due to the fact that they individually contain far less data than RS, the comparison between MT- and ST-SISSO is less insightful. We nonetheless report the result in the Supplementary Material.

\section{Conclusions}
In conclusion, we have introduced a nontrivial extension of the Sure Independence Screening and Sparsifying Operator (SISSO) algorithm. Such extensions is called Multi-Task (MT) SISSO, it belongs to the wider class of learning algorithms known as MT Learning, and is specifically designed for learning from databases with randomly or selectively distributed missing information. MT-SISSO finds a common descriptor, in terms of analytical functions of simple input physical quantities called {\em primary features}, when learning different properties (tasks) simultaneously. This joint learning yields robust models also with large amount of missing data, as demonstrated with two showcase materials-science examples: the prediction of the ground-state crystal structure for octet binaries compounds (out of 5 candidate structures) and the prediction of metal vs nonmetal classification of binary materials distributed over 17 crystal-structure prototypes. Since materials databases typically contain data from different sources and therefore unsystematic (different properties are collected for different materials), MT-SISSO is a method that can be suitably applied to these databases to yield predictive models for properties of interest.

The ST- and MT-SISSO package, as used for obtaining the results presented in this paper, is maintained by R. Ouyang and available open access at \texttt{github.com/rouyang2017/SISSO}.

\begin{acknowledgments}
This project has received funding from the European Union’s Horizon 2020 research and innovation program {(\#676580: The NOMAD Laboratory --- an European Center of Excellence and \#740233: TEC1p)}, the Berlin Big-Data Center (BBDC, \#01IS14013E), and  BiGmax, the Max Planck Society's Research Network on Big-Data-Driven Materials-Science.
\end{acknowledgments}

\newcommand{\Ozolins}{Ozoli\c{n}\v{s}}

\end{document}